\def\gtorder{\mathrel{\raise.3ex\hbox{$>$}\mkern-14mu
             \lower0.6ex\hbox{$\sim$}}}
\def\ltorder{\mathrel{\raise.3ex\hbox{$<$}\mkern-14mu
             \lower0.6ex\hbox{$\sim$}}}

\documentstyle[12pt,aaspp4]{article}
\begin{document}

\title{Detectability of High Redshift Ellipticals in the Hubble Deep Field}

\author{  Dan Maoz}

\affil{School of Physics \& Astronomy and Wise Observatory\\
Tel-Aviv University, Tel-Aviv 69978, Israel\\
dani@wise.tau.ac.il}

\begin{abstract}
Relatively 
few intensively star-forming 
galaxies at redshifts $z>2.5$ have been found in the Hubble Deep Field
(HDF). This has been 
interpreted to imply a low space density of elliptical galaxies
at high $z$, possibly 
due to a late ($z<2.5$) epoch of formation, 
or to dust obscuration of
the ellipticals that are forming at $z\sim 3$.
I use {\it HST}
UV ($\sim 2300$ \AA) images of 25 local early-type galaxies 
to investigate a third option, that ellipticals formed
at $z>4.5$, and were fading passively by $2<z<4.5$.
Present-day early-types are faint and centrally concentrated in the UV.
If ellipticals formed their stars
in a short burst at $z>4.5$, and have faded passively 
to their present brightnesses at UV wavelengths, they would
generally be below the HDF detection limits in any of its
bands at $z>2.5$. Quiescent $z\sim3$ ellipticals, if they exist,
should turn up in sufficiently deep IR images.

\end{abstract}


\section{Introduction}

Ellipticals are believed to be among the oldest galaxies,
with most of their stars having formed at early cosmic epochs,
and little star-forming activity since. 
(See Dickinson 1997, for a review of the evidence for this.) 
 Some arguments for a more
recent formation of ellipticals remain, however (e.g.,
 Kauffmann,
Charlot, \& White 1996).

An important step toward understanding the star-formation
history of the Universe was taken with the observation
of the Hubble Deep Field (HDF; Williams et al. 1996).
Based on earlier work by Guhathakurta et al. (1990),
Steidel \& Hamilton (1992), and Steidel, Pettini, \&
Hamilton (1995), Madau et al. (1996) have implemented
a method for identifying galaxies in the HDF at
redshifts $z\gtorder 2.5$. The cumulative effect
of H I in stellar atmospheres, in galaxies, and in
 absorption systems along the line of sight
causes galaxies to ``drop out'' of the F300W band
at $z\gtorder 2.5$, and out of the F450W band at 
$z\gtorder 3.5$, as the Lyman break passes through 
these bands. Steidel et al. (1996) have demonstrated,
by means of Keck spectroscopy of the brighter of the
HDF UV dropouts, the high efficiency of this method
in identifying high-$z$ galaxies. 

Madau et al. (1996) note that the high-$z$ galaxies in the HDF
are relatively faint. Their star-formation
rates are $<20 M_{\odot}$yr$^{-1}$, while rates in excess
of $50 - 100 M_{\odot}$yr$^{-1}$ would be expected from
ellipticals which were observed to be forming in a short $\sim 1$ Gyr
 burst at high $z$. They estimate a deficit by a factor
$\sim 10$ in  the comoving
 density of star-forming ellipticals at $<z>=2.75$,
 compared to the present-day density of $L>L_*$ ellipticals.
Assuming Madau et al.'s technique is efficient for the F450W
dropouts as well, a similar deficit is implied for the 
$3.5<z<4.5$ range.
This raises the possibility that, contrary to most evidence,
giant ellipticals (or at least the field population
among them)  formed at $z < 2.5$.

 Alternatively, Meurer et al. (1997) have argued that dust could
be dimming the light from actively-forming
 $z\sim 3$ early-type galaxies. They find that
the rest-frame UV colors of local starbursts and of the HDF
dropouts are similar. Since local starbursts are moderately
extinguished by dust, the same extinction could operate
in the high-$z$ galaxies. The extinction-corrected star formation
rate would then be consistent with the formation of ellipticals
at $z\sim 3$.

A third option, which I study here, 
is that ellipticals formed at even
higher $z$, and were already fading passively by $z\sim 3$.
The selection criteria used by Madau et al. (1996) are based
purely on colors in order to optimize the selection of high-$z$
{\it star-forming} galaxies. Quiescent, passively evolving  galaxies
are faint in the rest-frame UV, and hence
difficult to detect at high $z$. This has been realized since the
first space-UV observations of local galaxies and their use
for calculations of spectral $k$-corrections (e.g., Pence 1976; Coleman, Wu,
\& Weedman 1980; Bruzual 1983; King \& Ellis 1985; Wyse 1985; Kinney et al. 1996).
However, an accurate prediction of the appearance of high-$z$ galaxies
is complicated by the fact that galaxies are extended, structured, and
diverse objects, while spectral-$k$ corrections have always been based
on a small number of galaxies whose UV spectra were
 integrated over a set aperture.
In an attempt to obtain a ``morphological $k$-correction'',
Giavalisco et al. (1996)
have used UV images of seven (mostly spiral) 
nearby galaxies to 
 simulate the
appearance of galaxies in {\it Hubble Space Telescope (HST)} deep images.
They show that galaxies change their morphologies radically
when observed at 
high-$z$, due to the shift to the rest-frame UV.
(See also O'Connell \& Marcum 1996,
for a vivid demonstration of the effect.)
Giavalisco et al.
find that at $z>1.4$, most of their galaxies would be undetected
by {\it HST} in a 4.3 hour exposure in the F606W band. 

The simulations of Giavalisco et al. (1996) were restricted
to four normal spirals, one dusty starburst galaxy (M82), one Seyfert 2
galaxy (NGC 1068), and one dwarf elliptical (M32), that were
imaged with the {\it Ultraviolet Imaging Telescope (UIT)}.
 Furthermore, they simulated a
particular {\it HST} exposure time and band, and dealt with
the issue of galaxy fading in an approximate way.
Their work
 points to the need for a more
extensive and quantitative
examination of the expected appearance and brightness
of quiescent galaxies
at high $z$. Such an examination requires rest-frame UV images
of local galaxies spanning a range in optical properties,
particularly Hubble type and luminosity. 

Maoz et al. (1996b) carried out a UV (2300 \AA) imaging survey 
with the Faint Object Camera (FOC) on
{\it HST} of the central regions of 110 nearby galaxies.
The 110 galaxies are an unbiased selection (about one-half) from 
a complete sample of all large ($D_{25}>6'$) and nearby ($V<2000$ km s$^{-1}$)
galaxies. Since only the central $20''\times 20''$
were observed, the UV images of most of the galaxies in the sample 
are not useful for predicting the appearance of high-$z$ galaxies.
The early-type galaxies in the sample, however, generally have very faint,
diffuse, centrally-concentrated  UV emission, which becomes
undetectable in the FOC exposures already at a radii $<10''$. 
The projected physical radius probed by the FOC exposures at the distances
of these galaxies is typically $R\sim 1$ kpc. (The effective radius, $R_e$,
which includes half the light is $\sim 4$ kpc for an $L_*$ elliptical.)
Since the light profiles of ellipticals fall monotonically with radius
 in all bands,
it is unlikely that the UV emission rises again outside the 
FOC field of view. The FOC images of early-types can therefore
be used to estimate the appearance of the central and brightest
regions of high-$z$ galaxies of this kind.

In this {\it Letter} I examine the detectability of quiescent high-$z$
ellipticals in the HDF, based on {\it HST} UV observations
of 25 nearby early-type galaxies. The observed galaxies are all the 
early-types in the UV survey of Maoz et al. (1996b).
 The 25 galaxies span a range
of absolute luminosities and provide a representative
collection of local galaxies of this type. Apart from the
effects of redshift on their detectability, I also examine
the effect of passive evolution.

\section{Data}

Table 1 lists the 25 galaxies from Maoz et al. (1996b) that
have de Vaucouleurs T-type classifications $T\le 0$ (i.e.,
ellipticals and S0s). Also listed for each galaxy are its
heliocentric velocity, its optical
major and minor axis as listed in the ESO and UGC catalogs
(in tenths of arcminutes), its integrated B magnitude (from
de Vaucoulers et al. 1991), its absolute B magnitude
(see below), its Hubble type (from Sandage \&
Tammann 1987), and the total $f_{\lambda}(2300{\rm~\AA})$ 
in units  of $10^{-15}$ erg s$^{-1}$ cm$^{-2}$ \AA$^{-1}$, integrated 
above the background over the
entire $20''\times 20''$ area of the image, and $1\sigma$ uncertainty (from Maoz et al.
1996b). Although among the galaxies classified as early-types
by de Vaucoulers (1991) four are classified by Sandage \&
Tammann (1987) as Sa and one as Sb, such differences of one or two
Hubble classes are common when classifying galaxy-types by eye
(see Naim et al. 1995), and are not very meaningful.
Distances to these galaxies are generally not available, and are
difficult to estimate from their recession velocity alone because
they are so nearby. To get an idea of the distribution in absolute
luminosity of the sample, 
I have taken the distances estimated 
by Tully (1988) based mostly on recession velocity corrected for local flows.
For NGC 1023 and NGC 4649 I have used the distances measured
by Ciardullo, Jacoby, \& Tonry (1991) using planetary nebulae and for NGC 4486 (M87)
the surface brightness fluctuation distance found by Tonry et al. (1997).
 From the absolute magnitude entries in Table 1, the early-types
in the sample span a range from about $0.01L_*$ to $1.5L_*$, with
about half of the galaxies being approximately $L_*$ ($M_B=-20.7$).

  Figure 1 shows, for a selection of 12 of the galaxies,
 the FOC UV image from Maoz et al. (1996b) and an
optical image from the Palomar Sky Survey.
Note how, despite the fact that most of the galaxies are optically
bright, they are generally faint and centrally concentrated in the UV.
However, UV  emission {\it is} detected at the nucleus,
and measured with typically 5-sigma accuracy, when integrated over
the field of view (see Table 1). The flux from the central region
of such a galaxy, when observed at high $z$, can therefore be reliably calculated.
Since the light distribution in early types falls radially at all
bands (including the UV, as evidenced by these images), all the
surrounding regions of the galaxy will
necessarily have lower surface brightness
than the nuclear region probed by the FOC.

The FOC field of view is comparable to the {\it International Ultraviolet 
Explorer (IUE)} spectral aperture.
Meurer et al. (1995) and
Maoz et al. (1996) have shown
that there is generally an excellent agreement between nuclear UV fluxes
measured with the FOC and with {\it IUE}. One may wonder, then, what advantage
the present data have over existing spectra of ellipticals for
characterizing the broad-band UV flux of their central regions. First,
the morphological information in the FOC images allow one to see
what is in the aperture. For example, the FOC image of NGC 5253
(see Fig. 1) immediately reveals that it is an atypical
galaxy with a nuclear starburst, whereas someone simply collecting
all {\it IUE} spectra of ellipticals would include this galaxy as well.
Second, atlases and averages of {\it IUE} spectra (e.g.
Kinney et al. 1996) are collections of the objects that
happened to be observed by {\it IUE}. These collections are not complete samples in
any sense, and are dominated by the objects with the highest
S/N, i.e., the brightest ones. The average {\it IUE} elliptical spectrum
may therefore be biased by unrepresentatively blue ellipticals.
The galaxies in the FOC sample are an unbiased
selection from a complete sample of galaxies. This difference
is especially important in the present context, where we are trying
to study the UV detectability of typical galaxies.

\section{Moving the Galaxies to High Redshift}

I now calculate the appearance of the FOC early-type galaxies
if they were seen at various redshifts in the HDF bands.
The WF CCDs on WFPC2 have a scale of $0.1''$ pixel$^{-1}$, vs. 
$0.0225''$ pixel$^{-1}$ in the FOC in the zoomed $f/96$ format
used by Maoz et al. (1996b), i.e., a factor of 4.4. A given part of
the galaxy, when viewed at redshift $z$, would subtend an angle
on the sky given by its size divided by the angular diameter distance
(see, e.g., Giavalisco et al. 1996). Assuming a Virgo-like distance
of 15 Mpc for the typical galaxy in the UV sample, and
$H_0=70$ km s$^{-1}$ Mpc$^{-1}$, at redshift
$z=2.5$ the same region of the galaxy will subtend an angular extent 90 to
140 times smaller, for $q_0=1/2$ and $q_0=0.05$, respectively. The 
entire $1024\times 1024$ pixel FOC field of view therefore corresponds
to about $2\times 2$ WF pixels at $z> 2.5$. Fortuitously, this
area equals the minimum source detection area used by Williams
et al. (1996) to create the HDF source catalog.
Since their object detection algorithm 
searches for peaks
in such segments of the image and then ``grows'' them, an elliptical galaxy
whose nuclear region is undetected will remain undetected.
Therefore, despite the limited angular extent of the FOC
images of these large galaxies, the images can
 constrain the detectability of such galaxies at high $z$.

Table 1 lists the total observed flux density at 2300 \AA, $f_{\lambda}$(2300 \AA),
for each of the early-type galaxies, as measured by Maoz et al. (1996b).
The flux density of a galaxy at redshift $z$ is proportional to $D_L^{-2}(1+z)^{-1}$,
where $D_L$ is the luminosity distance. The central wavelengths of
the HDF bands, F300W, F450W, F606W, and F814W, correspond to 2300 \AA\ at 
redshifts of about 0.3, 1, 1.6, and 2.5, respectively. The flux density of
a galaxy in Virgo will decrease by a factor of $4\times 10^6$ ($1\times 10^7$) when it is moved
to $z=2.5$ for $q_0=1/2$ ($q_0=0$). Since the F300W dropouts in the HDF
are at $z\gtorder 2.5$, they are seen in their restframe UV even in the
longest wavelength (F814W) band.

 From Table 1 we see that the three early-type galaxies that are brightest
in the UV are NGC 3077, 5102, and 5253. From Figure 1, however, we see that
these galaxies are morphologically unusual, in that they display knots of
bright UV emission (NGC 3077, not shown in Figure 1, is similar in appearance
to NGC 5253). The UV emission probably extends outside the small FOC field of
view, as well. These knots of ``super star clusters'' turn up in large numbers
in {\it HST} UV images 
of galaxies where intense star formation is occuring (e.g.
Conti \& Vacca 1994; Meurer et al. 1995; Maoz et al. 1996a). In fact,
NGC 5102 and 5253 are well known starburst galaxies. All the other
early-type galaxies in the sample,
 however, are an order of magnitude fainter in their
UV images. They generally show only diffuse, 
centrally concentrated emission,
and sometimes some other faint features, such as a central
 compact source (e.g.,
NGC 404, NGC 4976) a jet (NGC 4486 = M87) or a circumnuclear ring (NGC 1079).
The brightest one among these more typical, quiescent, galaxies, NGC 4649, has
$f_{\lambda}(2300{\rm~\AA})= 
4.5\times 10^{-15}$ erg s$^{-1}$ cm$^{-2}$ \AA$^{-1}$, integrated over the
FOC field of view. At $z=2.5$, it would have
$f_{\lambda}(2300{\rm~\AA})= (4-10)\times 
10^{-22}$ erg s$^{-1}$ cm$^{-2}$ \AA$^{-1}$, depending on $q_0$.
In the AB magnitude system ($m_{AB}=-48.6 -2.5 \log f_{\nu}$, with
$f_{\nu}$ in units of erg s$^{-1}$ cm$^{-2}$ Hz$^{-1}$),
 this corresponds to an F814W magnitude
of 30.6 to 31.6 mag. From examination of the
Williams et al. (1996) object catalog, the $5\sigma$ limiting magnitude of the HDF in
this band for a 0.04 square arcsecond aperture ($2\times 2$ WF pixels)
 is about 29.2 mag,
 i.e., 1.5 to 2.5 mags brighter. At shorter wavelengths,
the UV flux from early-type galaxies is similar or less (e.g., Kinney et al.
1996), and the HDF detection
limits in the shorter wavelength-bands are similar.
The central $2\times 2$ WF pixels would therefore not be detected.
Since this is the brightest piece of the galaxy, the entire galaxy
would remain undetected.
 Hence {\it unevolving}
high-$z$ quiescent early-type galaxies are undetectable in the HDF.

However, even if ellipticals formed all their stars in a short 
burst at high $z$, some passive evolution in their luminosity
may be expected due to the aging of the stellar population. (See,
however, Driver et al. 1996, who argue that the fading caused by
the aging is, on average,
 balanced by the brightening caused by ongoing merging,
so that ``no evolution'' is a good approximation to the expected behavior.)
This evolution can be modeled by stellar population synthesis calculations.
Maoz et al. (1996a) showed that the 2300 \AA\ luminosity 
of an exponentially decaying starburst, after a few e-folding times,
can be parametrized with respect to time as $L_{2300}\propto t^{-1.4}$.
Spinrad et al. (1997) show that at 2300 \AA\ most of the light from
a co-eval stellar population is produced by stars at the main
sequence turnoff. Since much of the uncertainty in population
synthesis models arises from the different treatments of evolved stars,
the time dependence of the broad-band UV-luminosity  should be a robust
prediction of the models. The age of the Universe depends on redshift as
$t=1/H_0 (1+z)^{-\alpha}$, with $\alpha=1$ for $q_0=0$ to $\alpha=3/2$
for $q_0=1/2$. For $H_0=70$ km s$^{-1}$ Mpc$^{-1}$, the Universe at $z=2.5$ is 2 to 4 Gyrs
old for $q_0=1/2$ to 0. If, for example, ellipticals formed their stars
 in an early burst lasting 500 Myr, by $z=2.5$ active star formation
would have mostly ceased, and they would be fading passively with time.
Quiescent ellipticals at redshift $z$ will therefore be more luminous
than their present-day  counterparts
at rest-wavelength 2300 \AA\ by $(1+z)^{1.4 \alpha}$. At $z=2.5$,
this is a factor of 6 ($q_0=0$) to 16 ($q_0=1/2$), or 2 to 3 mags.

The brightest quiescent elliptical in the sample would therefore
just pass the HDF F814W magnitude limit, assuming a $q_0=1/2$
cosmology, and ignoring the opposite effect of merging on the
luminosity of the typical elliptical. Since these are extreme
assumptions, and most of the early-types
in the sample are about one magnitude fainter, I conclude that 
{\it most high-$z$ ellipticals that formed their stars in an early burst,
ending before $z>4.5$, are undetectable in the HDF.} An early epoch of
elliptical galaxy formation is therefore consistent with the
HDF results, and is a viable alternative to obscured star formation
at $z\sim3$ (Meurer et al. 1997) or late formation at $z< 2.5$ 
(Madau et al. 1996).

It is instructive to compare this result, based on the FOC data, 
to what one obtains using traditional $k$ corrections.
The elliptical galaxy template of Kinney et al. (1996), which
is based on four ellipticals observed by {\it IUE}, has
 an $f_{\lambda}(8140{\rm \AA})/f_{\lambda}(2300{\rm \AA})$ ratio of 22.
An $L^*$ elliptical with $M_B=-20.7$ has
an F814W AB magnitude of $-22.4$. At $z=2.5$, the integrated F814W
magnitude would be 28.3 ($q_0=1/2, H_0=70$ km s$^{-1}$ Mpc$^{-1}$, as above). 
If taking only the central 2 kpc diameter (corresponding to the 
FOC field of view of the local galaxies) of a de Vaucouleurs
profile with $R_e=4$ kpc,
the observed magnitude would be 30.2, vs. 30.6 derived from the
FOC data. The agreement is reasonable, but the spectral $k$ correction
overpredicts the brightness of high-$z$ ellipticals. Furthermore,
recall that the galaxy used in this calculation, NGC 4649, is
the brightest one, rather than the representative one,
 in the UV among the quiescent $\sim L_*$ early-types in
the FOC sample. 

Part of the UV flux in the FOC images may be emission
from some residual star formation, or
``UV-upturn'' emission, generally thought to originate in
 evolved stellar populations (e.g. Dorman, O'Connell, \& Rood 1995).
 In that case,
however, only a fraction of the present-day UV emission is
 from the main-sequence population, i.e., the main sequence UV emission is
even smaller than observed. This will only strengthen
 the conclusion on the undetectability of the high-$z$ ellipticals
that are viewed in their rest-frame UV.

It may be argued that the distances of the galaxies in the
sample are uncertain, compromising the accuracy of
 the luminosities and the predicted HDF
magnitudes. However, seven of the FOC early-types, all fairly bright,
are in Virgo and/or have accurate distances, so their
luminosities are no more uncertain than that of most galaxies.
Among the 10 most luminous ($M_B>-20$)
galaxies, on which the argument hinges, three are in Virgo (two
with accurate distances), and four others have velocities $>1500$ km s$^{-1}$,
so their Hubble distances are probably already not that bad an
approximation. It is unlikely that all of these galaxies are 
anomalously closer than it seems, and hence substantially sub-$L_*$, or that their
distances are all much beyond 15 Mpc, and that this is the cause
of their UV faintness. By setting an upper limit on the expected brightness
of high-$z$ ellipticals which is based on the {\it brightest}
 quiescent galaxy in the local sample, the conclusion is robust to
the uncertainty in distances.

Can the detectability of ellipticals in the HDF be improved by using
a larger detection aperture? Assuming a background-noise-limited
detection, and that the UV light follows a de Vaucouleurs surface
brightness profile, 
the signal-to-noise ratio as a function of the aperture radius $R$ is
\begin{equation}
\frac{S}{N}(x)\propto \frac{\int_0^x x e^{-7.67x^{1/4}}dx}{x},
\end{equation}
where $x=R/R_e$.
The optimal aperture for detection is at the maximum of this function, which
occurs at $x\approx 0.14$. Since the FOC images typically extend to $x=0.25$,
using a larger aperture in the HDF will not change the conclusions, unless
the UV light distribution falls off much less steeply than a de Vaucouleurs
profile.
This can be examined in the future using UV images of local galaxies having
a larger field of view than the FOC.

As is well known (see, e.g., Ellis 1997), near-infrared (IR) $k$ corrections
are much smaller than optical ones, so high-$z$ ellipticals
may be more readily detected in the IR. In the context 
of the present discussion,
the $B$ and $V$ spectral regions are shifted into
the infrared $H$ ($1.65 \mu m$) and $K$ ($2.2 \mu m$) bands,
 respectively, at
$z=2.6\sim 2.8$. The integrated brightness of an
$\sim L_*$ elliptical is $V\sim 11.0$ mag at Virgo, and the
brightness of the fraction of the galaxy within an
aperture of radius $0.25R_e$ (a $\sim 0.25''$ diameter aperture
at $z>2.5$) is $V\sim 12.8$. The galaxy flux within this aperture
will be $2-5 \times 10^{-21}$ erg s$^{-1}$ cm$^{-2}$ \AA$^{-1}$
in the $K$ band at $z=2.7$, depending on $q_0$. In $f_{\nu}$ units, this
is $3-8\times 10^{-31}$ erg s$^{-1}$ cm$^{-2}$ Hz$^{-1}$,
or $0.03-0.08$ $\mu$Jy, or a Johnson $K$ magnitude of 25--26 mag.
As at UV wavelengths, passive evolution can potentially make
the galaxy several magnitudes brighter at high $z$.
Such surface brightnesses are challenging, but detectable
in long exposures with NICMOS on {\it HST} or on large
ground-based telescopes. 
 If passively fading ellipticals
existed at these redshifts, they will turn up
in deep enough IR images. Objects of these brightnesses
that are undetected in optical images of the same field
(e.g. the HDF) would be prime candidates for passively evolving
ellipticals that formed at high-$z$. Apart from revealing
the formation history of ellipticals,
the  age estimation
of such galaxies, if detected, can set powerful constraints
on $q_0$ (as in Spinrad et al. 1997). Alternatively, the
non-detection of such a population in the IR would indicate
that field ellipticals formed at $z\sim 3$ but were obscured,
or have formed more recently than $z=2.5$.

\acknowledgements
I thank Piero Madau, Gerhardt Meurer, Amiel Sternberg, and the anonymous referee
 for valuable comments.
This work was supported by grant 94-00300 from the U.S.-Israel
Binational Science Foundation and by a grant from the Israel
Science Foundation.

								     

\newpage
\begin{deluxetable}{lrrrrrrlcc}
\footnotesize
\tablewidth{0pt}
\tablecaption{Early-Type Galaxies in the FOC UV Survey}
\tablehead{\colhead {NGC} & 
\colhead {$V_h$} & \colhead {$D_a$} & \colhead
 {$D_b$} &
\colhead {$B_{\rm mag}$} &\colhead{$M_B$}& \colhead {T}&\colhead{Classif.}&
\colhead{$f_{UV}$}
&\colhead{$\sigma$}\nl
\colhead {(1)}&\colhead {(2)}&\colhead {(3)}&\colhead {(4)}&
\colhead {(5)}&\colhead {(6)}&\colhead {(7)}&\colhead {(8)}&\colhead {(9)}&\colhead{(10)}
\nl}
\startdata
 185 &$-227 $&$144$&$120$&$11.00$&$-$13.2&$-5$&$dE3 pec  $&$ 1.5$&$ 0.4$\nl   
 404 &$ -36 $&$ 60$&$ 60$&$11.30$&$-$15.7&$-3$&$S0_3     $&$ 3.5$&$ 0.2$\nl  
1023 &$ 648 $&$ 85$&$ 40$&$10.50$&$-$19.5&$-3$&$SB0_1    $&$ 4.0$&$ 1.1$\nl  
1079 &$1465 $&$ 70$&$ 50$&$12.30$&$-$18.8&$ 0$&$Sa       $&$ 2.8$&$ 0.2$\nl
1291 &$ 839 $&$130$&$130$&$ 9.42$&$-$20.3&$-3$&$SBa      $&$ 2.9$&$ 0.3$\nl
1332 &$1469 $&$ 60$&$ 20$&$11.20$&$-$20.0&$-3$&$S0_1     $&$ 2.9$&$ 0.2$\nl
1543 &$1088 $&$ 70$&$ 70$&$11.57$&$-$19.1&$-2$&$RSB0/a   $&$ 1.7$&$ 0.2$\nl
2768 &$1363 $&$ 65$&$ 30$&$11.10$&$-$20.8&$-5$&$S0_{1/2} $&$ 1.2$&$ 0.4$\nl  
2784 &$ 708 $&$ 90$&$ 50$&$11.25$&$-$18.0&$-2$&$S0_1     $&$ 1.5$&$ 0.2$\nl 
3077 &$   7 $&$ 60$&$ 45$&$10.70$&$-$15.9&$ 0$& \nodata   &$21.9$&$ 5.0$\nl  
4438 &$  86 $&$ 97$&$ 39$&$12.00$&$-$19.1&$ 0$&$Sb       $&$ 1.5$&$ 0.3$\nl  
4486 &$1292 $&$ 70$&$ 70$&$10.40$&$-$20.6&$-6$&$E0       $&$ 1.6$&$ 0.2$\nl
4636 &$ 937 $&$ 70$&$ 50$&$11.80$&$-$19.3&$-5$&$E/S0_1   $&$ 1.8$&$ 0.3$\nl  
4649 &$1095 $&$ 70$&$ 60$&$10.30$&$-$20.6&$-5$&$S0_1     $&$ 4.5$&$ 0.4$\nl  
4762 &$1006 $&$ 90$&$ 20$&$11.10$&$-$20.0&$-2$&$S0_1     $&$ 2.0$&$ 0.4$\nl  
4866 &$1980 $&$ 60$&$ 13$&$11.90$&$-$19.1&$-1$&$Sa       $&$ 1.2$&$ 0.2$\nl 
4976 &$1503 $&$ 60$&$ 35$&$11.17$&$-$20.2&$-5$&$S0_1     $&$ 2.5$&$ 0.6$\nl
5023 &$ 400 $&$ 75$&$  8$&$13.20$&$-$15.7&$-5$& \nodata   &$ 2.2$&$ 0.4$\nl  
5084 &$1739 $&$190$&$ 28$&$12.02$&$-$20.1&$-2$&$S0_1     $&$ 1.2$&$ 0.3$\nl
5101 &$1864 $&$ 70$&$ 70$&$11.52$&$-$20.7&$ 0$&$SBa      $&$ 1.5$&$ 0.4$\nl
5102 &$ 420 $&$120$&$ 50$&$10.35$&$-$17.5&$-3$&$S0_1     $&$20.2$&$ 1.3$\nl
5195 &$ 558 $&$ 70$&$ 50$&$10.60$&$-$19.3&$ 0$&$SB0_1 pec$&$ 1.5$&$ 0.4$\nl  
5253 &$ 417 $&$ 60$&$ 20$&$10.99$&$-$16.5&$ 0$&$Amorph.  $&$95.4$&$ 6.2$\nl
5322 &$1804 $&$ 60$&$ 40$&$11.30$&$-$21.2&$-5$&$E4       $&$ 3.2$&$ 0.8$\nl  
5866 &$ 672 $&$ 65$&$ 30$&$11.10$&$-$19.8&$-1$&$S0_3     $&$ 1.5$&$ 0.4$\nl
\enddata							     
\tablecomments{ (2) heliocentric velocity, in km s$^{-1}$;
(3)(4) optical major and minor axis as listed in the ESO and UGC catalogs
(in tenths of arcminutes); (5) integrated B magnitude (from
de Vaucoulers et al. 1991); (6) absolute $B$ magnitude, assuming
approximate distanced from Tully (1988); (7)
de Vaucouleurs T-type classification; (8) Hubble type (from Sandage \&
Tammann 1987); (9)(10) total $f_{\lambda}(2300{\rm~\AA})$ 
in units  of $10^{-15}$ erg s$^{-1}$ cm$^{-2}$ \AA$^{-1}$, integrated 
above the background over the
$20''\times 20''$ area of the image, and $1\sigma$ uncertainty (see Maoz et al.
1996b).}
\end{deluxetable}						     

\begin{figure}
\epsscale{1.0}
\plotone{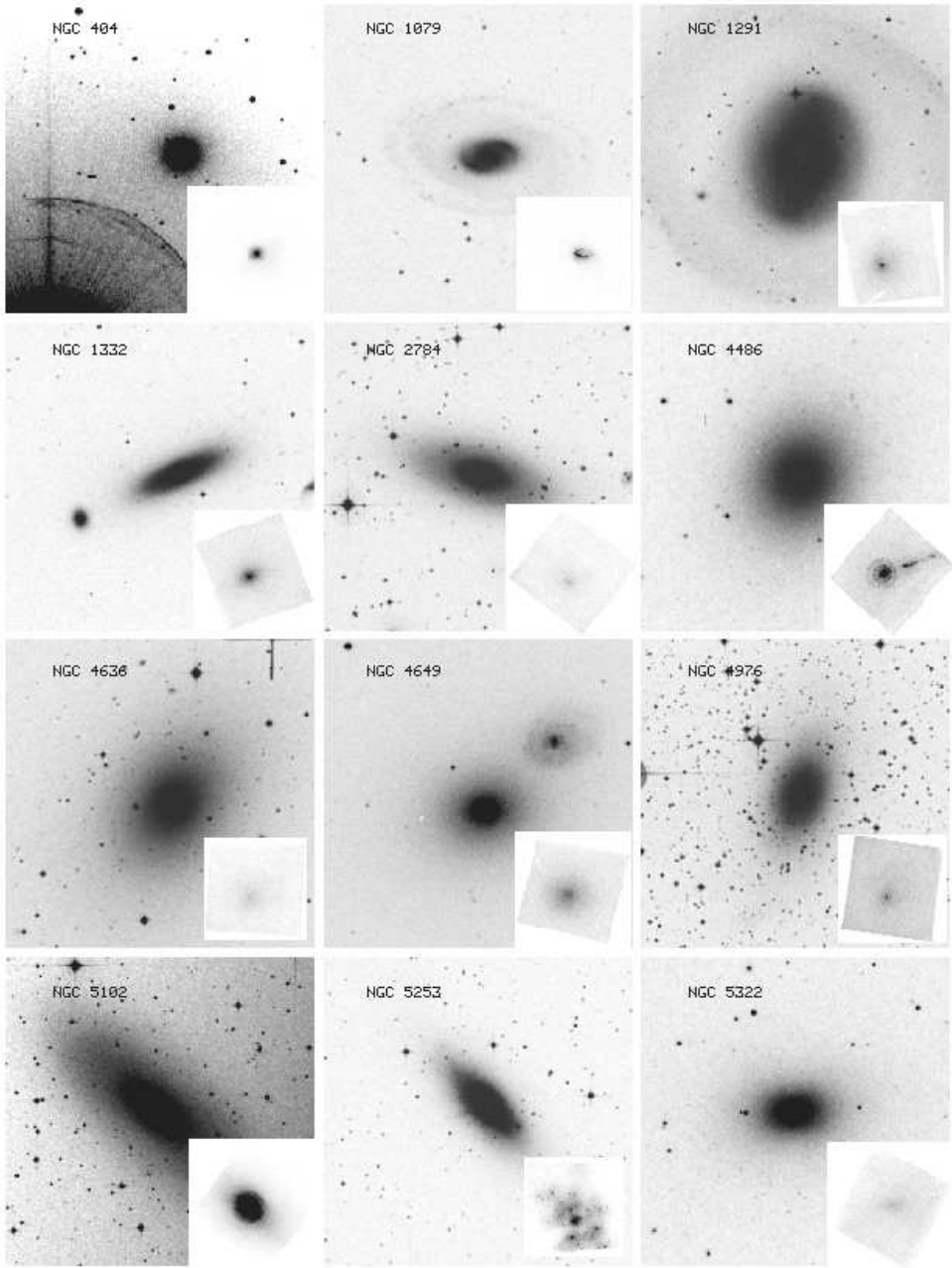}
\caption{Visual-band images of a selection of early-type galaxies
and (inset) {\it HST} FOC 2300 \AA\ images of their centers. The 
visual images are from the Palomar Sky Survey and measure $8'$ on a 
side. The field of view of the UV images is $22''\times 22''$ 
(except for NGC 4486 and NGC 5102, whose fields are $10''\times 10''$).
They have been rotated so that north is up and east to the left,
as in the visual-band images. Note the feeble and concentrated
UV emission from the central regions of most of
 these optically-bright galaxies.}
\end{figure}
								     
\end{document}